# Electrically switchable $2^N$-channel wave-front control with $N$ cascaded polarization-dependent metasurfaces

Zhiyao Ma, Tian Tian, Yuxuan Liao, Xue Feng*, Yongzhuo Li, Kaiyu Cui, Fang Liu, Hao Sun, Wei Zhang and Yidong Huang*

Department of Electronic Engineering, Tsinghua University, Beijing 100084, China

*Corresponding author: x-feng@tsinghua.edu.cn (X.F.); yidonghuang@tsinghua.edu.cn (Y.H.)

**Abstract**

Metasurfaces with tunable functionalities are greatly desired for modern optical system and various applications. To increase the operating channels of polarization-multiplexed metasurfaces, we proposed a structure of $N$ cascaded dual-channel metasurfaces to achieve $2^N$ electrically switchable functional channels without intrinsic noise or cross-talk. As proof of principles, we have implemented a 3-layer setup to achieve 8 channels. In success, we have demonstrated two typical functionalities of vortex beam generation with switchable topological charge of $l$=-3 ~ +4 or $l$=-1~ -8, and beam steering with the deflecting direction switchable in an 8×1 line or a 4×2 grid. We believe that our proposal would provide a practical way to significantly increase the scalability and extend the functionality of polarization-multiplexed metasurfaces, which are potential for the applications of LiDAR, glasses-free 3D display, OAM (de)multiplexing, and varifocal meta-lens.

**Introduction**

Metasurfaces[1,2] arouse broad research interest due to the unprecedented manipulation of lightwave in terms of phase[1,2], amplitude[3–5], polarization[6,7], and frequency[8,9]. Metasurfaces possess rich advantages over traditional planar optical devices, including compact footprint and thickness[10,11], subwavelength-resolution control of lightwave[7,12,13], low-loss tramsmition[14] as well as versatile functionalities[15,16]. In particular, with the abrupt phase-shifting ability in subwavelength scale[1], metasurfaces have gained significant progress in the fixed wave-front controlling, *e.g.* beam steering[17,18], meta-lens[19,20], vortex beam generat[4,6,21], holograms[22,23], *etc.*. With the growing need of optical systems, metasurfaces with tunable functionalities are further concerned. One approach is dynamically reconfiguring the subwavelength structure by various mechanisms, including electro-optic[24–27], mechanical[28–30], phase transition[31,32], *etc.*. Thus, the response of the metasurface can be tuned with reconfiguring structure. However, the applicable functionalities with most mechanisms are quite limited, and the extra complexity required for active reconfiguring would significantly affect the scalability. Till now, such reconfigurable metasurfaces remain technically challenging. Another approach depends on the states of input lightwave, which serve as knobs for tuning the response of fixed metasurfaces[33], thus tuning the functionalities in success. The properties include incident angle[34,35] or direction[36,37], wavelength[38–41], carried orbital angular momentum (OAM)[42,43]. Besides, with two cascaded metasurfaces, the functionalities can be also tuned by the relative displacement[44–46] and rotation[47,48]. Actually, with these methods, the input lightwave at the second metasurface is manipulated equivalently, while the structures of both metasurfaces are fixed.

As an intrinsic property of lightwave, polarization has been used for tuning the response of metasurfaces as well, which is also regarded as polarization multiplexing[49–53]. Polarization multiplexing can be achieved by employing polarization-dependent subwavelength structures, without compromising on resolution and information capacity of metasurfaces[52]. For such polarization-dependent metasurfaces, the functionalities can be independently designed for two orthogonal polarization states of input light, *e.g.* horizontal and vertical linear polarization[51]. As for specific functionalities, dual-channel lenses[49,54], dual-channel holograms[51], and dual-channel vortex beam generators[50] have been achieved. Since the DoFs of polarization state is only two, the number of independent channels is also limited to two for a single metasurface. This limitation can be broken by employing a polarizer after the metasurface, so that the off-diagonal elements of the Jones matrix can be extracted as the third independent channel[55]. Furthermore, engineered noise have been employed to increase the number of channels up to 11 with moderate cross-talk[56]. These approaches significantly increased the scalability of polarization-multiplexed metasurfaces. However, there would be intrinsic loss and additional design complexity because the required Jones matrices are non-unitary[55,56]. Besides, although the introduced cross-talk is moderate in holograms, it could be unacceptable in noise-sensitive or cross-talk-sensitive applications, *e.g.* OAM (de)multiplexing, LiDAR. Generally, the independent channels without intrinsic cross-talk are still limited to three. Therefore, it is still challenging for further increasing channels and extending the functionality of polarization-multiplexed metasurfaces.

In this paper, we propose a practical approach to increase the number of channels to $2^N$ without intrinsic noise or cross-talk by cascading $N$ metasurfaces. To increase the DoFs on polarization, a direct consideration is to cascade multiple polarization-dependent metasurfaces. Cascading $N$ metasurfaces combined with polarization controllers would increase the total DoFs on polarization up to $N+1$, which would be further discussed in the section of Results. Obviously, the cost of cascading $N$ metasurfaces would be too high if the number of the obtained channels is only $N+1$. However, we found that the channels can be increase to $2^N$ under certain condition. The condition is that the phase pattern of cascaded structure should have the same function form with that of a single metasurface, while the parameters of the cascaded structure are summation of each single metasurface. Such condition could be satisfied in most cases, including beam steering, vortex beam generation, lens and so on. Thus, the channels of polarization-multiplexing can be increased exponentially for most common dynamic wavefront-controlled functionalities and applications.

As proof of principle, we have implemented an 8-channel setup through cascading 3 layers of metasurfaces. Thus, 8-channel vortex beam generation as well as beam steering have been experimentally demonstrated and characterized, respectively. In vortex beam generation, the topological charge (TC, denoted as *l*) can be switched within the range of *l*=-3 ~ +4 or *l*=-1~ -8. While in beam steering, the deflecting direction can be switched in an 8×1 line or a 4×2 grid, and the transmittance is within the range from 15% to 25%. Besides, since the employed polarization controllers are electrically tunable liquid crystal (LC) phase retarders, the structure can be electrically switched among different channels.

# Results
## Pinciple of $2^N$-channel switching

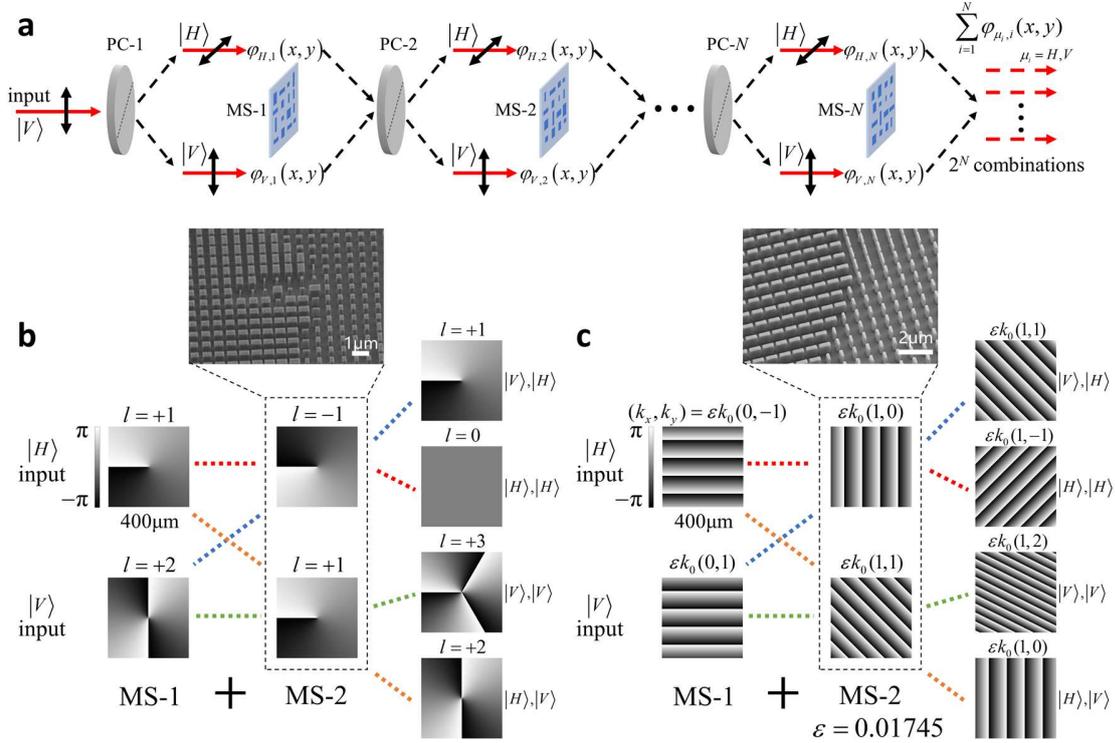

Fig. 1 Principle of $2^N$-channel switchable wave-front control. (**a**) The schematic of the cascaded structure. $N$ polarization-dependent dual-channel metasurfaces are cascaded. The initial input state is $|V\rangle$. A polarization controller (PC) is set before each metasurface to switch the input state at each metasurface between $|H\rangle$ and $|V\rangle$, thus one of the dual phase patterns are picked up. The whole phase pattern would be the sum of all picked phase patterns. (**b**) Example of cascading two dual-channel vortex phase patterns to generate four channels of vortex phase patterns. The TCs of the first pattern are $l$=+1, +2 for $|H\rangle$ and $|V\rangle$ input, respectively. While the TCs of the second pattern are $l$=-1, +1, with the inset of a SEM image of the corresponding metasurface. The TCs of four cascaded channels are $l$=+1, 0, +3, +2. (**c**) Example of cascading two dual-channel blazed gratings to generate four channels of blazed gratings. The momenta of the first pattern are $(k_x,k_y)$=$\varepsilon k_0(0,-1)$, $\varepsilon k_0(0,1)$ for $|H\rangle$ and $|V\rangle$ input, respectively. While the TCs of the second pattern are $(k_x,k_y)$=$\varepsilon k_0(1,0),\varepsilon k_0(1,1)$, with the inset of a SEM image of the corresponding metasurface. The momenta of four cascaded channels are $(k_x,k_y)$=$\varepsilon k_0(1,1),\varepsilon k_0(1,-1),\varepsilon k_0(1,2),\varepsilon k_0(1,0)$. $\varepsilon$=0.01745.

Based on a polarization-dependent metasurface, two independent phase-only patterns can be designed to modulate on the wave-front of the input lightwave with two orthogonal polarization states respectively. Such two phase patterns correspond to two functional channels, thus dual-channel switchable wavefront-control can be achieved. Since the DoFs of polarization state is only 2, there are only two independent functional channels of a single polarization-dependent metasurface in theory. To increase the number of channels, our proposal is to cascade $N$ dual-channel metasurfaces, while setting $N$ polarization controllers before each metasurface as shown in Fig. 1a. Without loss of generality, suppose the pre-settled orthogonal polarization states of input light of each metasurface are horizontal and vertical linear polarization, noted $|H\rangle$ and $|V\rangle$. It should be mentioned that the employed orthogonal states can be another set, *e.g.* left-hand and right-hand circular polarization (Ref). The polarization controllers are employed to switch the polarization state of input light at each metasurface between $|H\rangle$ and $|V\rangle$. As a result, for each metasurface, one of the two designed phase patterns is picked out. Thus, in the whole structure, there will be $2^N$ combinations of cascaded phase patterns. Suppose the dual phase patterns of the *i*-th metasurface are $\varphi_{H,i}(x,y)$ and $\varphi_{V,i}(x,y)$, the equivalent phase pattern of the whole structure can be expressed by the summation of the selected phase pattern from each metasurface:

$$\varphi_\mu(x,y) = \sum_{i=1}^{N} \varphi_{\mu_i,i}(x,y)$$

$$\begin{bmatrix} \varphi_{[H,H,\ldots,H]^T} \\ \varphi_{[V,H,\ldots,H]^T} \\ \vdots \\ \varphi_{[V,V,\ldots,V]^T} \end{bmatrix}_{2^N \times 1} = \begin{bmatrix} 1 & 0 & 1 & 0 & \cdots & 1 & 0 \\ 0 & 1 & 1 & 0 & \cdots & 1 & 0 \\ \vdots & \vdots & \vdots & \vdots & \ddots & \vdots & \vdots \\ 0 & 1 & 0 & 1 & \cdots & 0 & 1 \end{bmatrix}_{2^N \times 2N} \begin{bmatrix} \varphi_{H,1} \\ \varphi_{V,1} \\ \varphi_{H,2} \\ \varphi_{V,2} \\ \vdots \\ \varphi_{H,N} \\ \varphi_{V,N} \end{bmatrix}_{2N \times 1} \quad (1)$$

Where $\mu$ is a $N \times 1$ vector representing input polarization states of each metasurface. Since $\mu_i$ is either $H$ or $V$ and determined by the *i*-th polarization controller, $\mu$ can take $2^N$ distinct values. For more clarity, the matrix-vector multiplication form is expanded in the second line of Eq. 1. It should be mentioned that in the coefficient matrix, the sum of the $2m-1$ column and the $2m$ column for any integer $m \leq N$ is always a column vector with all elements of 1, thus the rank of the matrix is actually $N+1$. That is to say, the number of independent channels is constrained to $N+1$ in general, although there are $2^N$ cascaded phase patterns.

Here, the concept of **independent** should be discussed in details. The number of combinations for the cascaded phase pattern is still $2^N$, but there would be correlation among them. Thus, these $2^N$ cascaded phase patterns are not independent, as well as the corresponding functionalities. In other words, only $N+1$ among the $2^N$ cascaded phase patterns can be arbitrarily or independently designed, while the others would be determined by them. This corollary is mentioned in the section of Introduction.

However, under certain conditions, it would be possible to achieve $2^N$ channels with distinct functionalities. The condition is that phase pattern of cascaded structure should have the same function form with that of a single metasurface, while the parameters of the cascaded structure are summation of each single metasurface. To be more specific, for most common functionalities, the target of wavefront-control is to manipulate certain parameters of the light field. For such a functionality, suppose the form of the phase pattern can be written as $\varphi_F(x, y, p^{(1)}, p^{(2)}, \ldots)$, which is a map of the target parameters $p^{(1)}, p^{(2)}, \ldots$. If we consider a functionality, in which $\varphi_F$ is a linear map of these target parameters, the equivalent phase pattern of multiple cascaded metasurfaces with such functionality can be rewritten as:

$$\sum_{i=1}^{N} \varphi_F(x, y, p_i^{(1)}, p_i^{(2)}, \ldots) = \varphi_F(x, y, \sum_{i=1}^{N} p_i^{(1)}, \sum_{i=1}^{N} p_i^{(2)}, \ldots) \quad (2)$$

As seen in Eq. 2, the equivalent phase pattern can be rewritten from cascaded phase patterns into a set of cascaded parameters $\sum_{i=1}^{N} p_i^{(1)}, \sum_{i=1}^{N} p_i^{(2)}, \ldots$, which is still in the form of $\varphi_F$.

Thus, the equivalent functionality of cascaded metasurfaces would be as same as a single metasurface, while the parameters are modified.

It should be mentioned that the number of target parameters is determined by the functionality. Specifically, there are several examples of functionalities with such linear-map property. For the functionality of vortex beam generation, the required vortex phase pattern can be written as $\varphi_{OAM}(x,y,l) = l \cdot \arctan2(y,x)$ [57]. Obviously, the phase pattern is a linear map of a single target parameter that is the TC $l$. As seen in Eq. 3, cascading vortex phase patterns is equivalent to a vortex phase pattern as well. The cascaded TC would be the summation of TCs of each single pattern. For the functionality of beam steering, the required phase patterns are blazed gratings with $\varphi_{BG}(x,y,k_x,k_y) = k_x x + k_y y$ [58]. Here, the phase pattern is a linear map of two target parameters, which are the transverse momenta $k_x$ and $k_y$, respectively. As seen in Eq. 4, cascading blazed gratings is equivalent to a blazed grating as well. The cascaded momenta would be the sum of momenta of each single pattern.

$$\sum_{i=1}^{N} \varphi_{OAM}(x,y,l_i) = \sum_{i=1}^{N} (l_i \cdot \arctan2(y,x)) = \arctan2(y,x) \cdot \left( \sum_{i=1}^{N} l_i \right) \quad (3)$$

$$\sum_{i=1}^{N} \varphi_{BG}(x,y,k_{x,i},k_{y,i}) = \sum_{i=1}^{N} (k_{x,i} x + k_{y,i} y) = x \sum_{i=1}^{N} (k_{x,i}) + y \sum_{i=1}^{N} (k_{y,i}) \quad (4)$$

Besides, for the functionality of lens, the required phase pattern can be written as a linear-map of the reciprocal focal length[59] as well. However, the functionality of hologram is a counter-example, where the required phase pattern is numerically calculated and cannot be written as a linear map of explicit parameters. Thus, the equivalent functionality of multiple cascaded holograms cannot guarantee a meaningful hologram.

Combining the linear-map property (Eq. 2) with the cascaded polarization-dependent metasurfaces, $2^N$ functional channels can be achieved as shown in Fig. 1a. Here, the dual phase patterns of each single metasurface in Eq. 1 should be determined by the mapping relation $\varphi_F$ and corresponding target parameters. For the $i$-th polarization-dependent metasurface, suppose the parameters of the phase pattern for input polarization state $|H\rangle$ are $p_{H,i}^{(1)}, p_{H,i}^{(2)},...$, while the parameters for $|V\rangle$ are $p_{V,i}^{(1)}, p_{V,i}^{(2)},...$. Then the phase patterns of the $i$-th metasurface can be rewritten as:

$$\begin{aligned} \varphi_{H,i}(x,y) &= \varphi_F(x,y,p_{H,i}^{(1)},p_{V,i}^{(2)},...) \\ \varphi_{V,i}(x,y) &= \varphi_F(x,y,p_{H,i}^{(1)},p_{V,i}^{(2)},...) \end{aligned} \quad (5)$$

To write the cascaded phase patterns, substitute $\varphi_{H,i}$'s and $\varphi_{V,i}$'s in Eq. 1 with Eq. 5, and apply the linear-map property in Eq. 2:

$$\varphi_\mu(x,y) = \sum_{i=1}^{N} \varphi_F(x,y,p_{\mu_i,i}^{(1)},p_{\mu_i,i}^{(2)},...) = \varphi_F(x,y,\sum_{i=1}^{N} p_{\mu_i,i}^{(1)}, \sum_{i=1}^{N} p_{\mu_i,i}^{(2)},...) \quad (6)$$

As seen in Eq. 6, for $2^N$ combinations of input polarization states, all the $2^N$ equivalent phase patterns $\varphi_\mu$ would achieve the same functionality of $\varphi_F$, with $2^N$ sets of cascaded parameters. To conclude, $2^N$ distinct functional channels can be achieved. For a simple example of $N=2$, Fig. 1b shows cascading dual-channel vortex phase patterns generate four channels of vortex phase patterns. The TCs of the first pattern are $l=+1, +2$ for $|H\rangle$ and $|V\rangle$ input, respectively. While the TCs of the second pattern are $l=-1, +1$, with SEM image of the corresponding metasurface shown in the inset of Fig. 1b. The TCs of four cascaded channels are $l=+1, 0, +3, +2$. Another example of cascading two dual-channel blazed gratings to generate four channels of blazed gratings is shown in Fig. 1c. The momenta of the first pattern are $(k_x,k_y)=\varepsilon k_0(0,-1), \varepsilon k_0(0,1)$, where $\varepsilon=0.01745$. While the TCs of the second pattern are $(k_x,k_y)=\varepsilon k_0(1,0),\varepsilon k_0(1,1)$, with SEM image of the corresponding metasurface shown in the inset of Fig. 1c. The momenta of four cascaded channels are $(k_x,k_y)=\varepsilon k_0(1,1),\varepsilon k_0(1,-1),\varepsilon k_0(1,2),\varepsilon k_0(1,0)$.

It should be mentioned that there are no intrinsic cross-talk between channels, since the phase patterns are strictly with the form of $\varphi_F$. Besides, there are no intrinsic loss since the required Jones matrices of metasurfaces and polarization controllers are all unitary (details can be found in Supplementary Text 1).

**Experimental Setup**

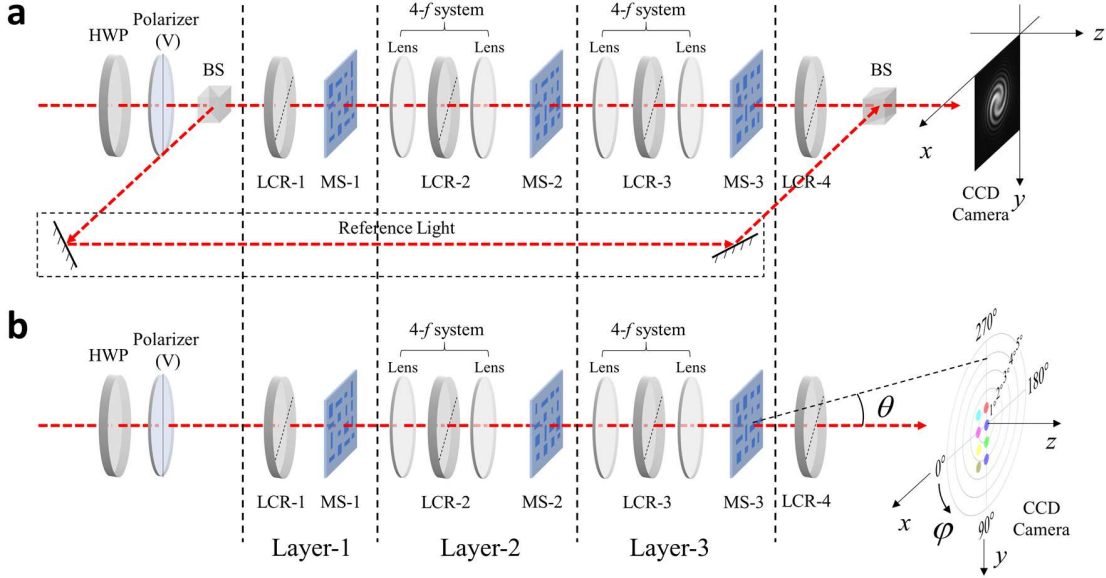

Fig. 2 Implementation of 8-channel switchable wave-front control with 3-layer setup. The input fundamental Gaussian beam is filtered into polarization state $|V\rangle$. Liquid crystal retarders (LCR) with slow axis at 45° are utilized as polarization controllers. The 4-$f$ systems are employed to guarantee the cascaded phase pattern is simply the sum of each single phase pattern, without considering the propagation between metasurfaces. The fourth LCR is employed to switch the output polarization state back to $|V\rangle$. (**a**) In the characterization of TCs of generated vortex beam, a path of reference light is employed for the interference fringes. An example of interference fringe with $l$=-2 is show on the right. (**b**) In the characterization of beam steering, the deflecting direction is characterized with $\theta$ and $\varphi$ in spherical coordinates. An example of distribution of 8 deflecting directions is show on the right.

To verify our proposal, a 3-layer setup is implemented for 8-channel switchable wave-front control. The functionality depends on the designed phase patterns of metasurfaces. Two functionalities of vortex beam generation and beam steering are demonstrated and characterized, respectively. Fig. 2a shows a schematic of the optical setup for characterization of vortex beam generation, while Fig. 2b shows that for beam steering. An additional reference light is required for the interference fringes to identify the TC of generated vortex beams, while the deflecting direction of beam steering would be characterized with $\theta$ and $\varphi$ in spherical coordinates. Details of the devices and the photograph of the optical setup can be found in Supplementary Text 2.

The metasurfaces are composed of rectangular nanopillar of amorphous silicon ($\alpha$-Si) on a SiO$_2$ substrate, and the operation wavelength is considered as 1550nm. The designed height of nanopillars is 900nm while the period of arrangement is 800nm, and the transverse size ranges from 120nm to 700nm. Each single metasurface is a square with side length ~640μm for vortex beam generation, while the side length is ~400μm for beam steering. Examples of the SEM image of the fabricated metasurfaces are presented in the inset of Fig. 1b and Fig. 1c. The design process of the metasurface can be found in Supplementary Text 3, and the fabrication process can be found in our previous work[60].

Here, the input polarization state can be either $|H\rangle$ or $|V\rangle$, while $|V\rangle$ is picked up in this work. Liquid-crystal (LC) phase retarders are utilized as the polarization controllers before each metasurface. To electrically switch the input states between $|H\rangle$ and $|V\rangle$, the slow axis of LC phase retarder is set at 45°. Therefore, for either $|H\rangle$ or $|V\rangle$ input on the LC phase retarder, the state can be varied from one to the other with the phase retardance of π, while the state would keep constant with the phase retardance of 0. By setting appropriate combination of the phase retardance on each LC phase retarder, the input polarization states can be switched among 8 combinations, then the equivalent phase pattern of the whole structure can be switched among 8 channels.

Besides the 3 LC retarders before each metasurface, another LC retarder is settled after the third metasurface so that the output polarization state of each channel is switched back to $|V\rangle$. Finally, the far-field intensity is captured by a CCD camera.

**Experimental results of 8-Channel vortex beam generation**

For the vortex beam generation, there is only one target parameter of the TC *l*. Here, each single metasurface is design as a dual-channel vortex beam generator with the phase pattern form of Eq. 3. It should be mentioned that each phase pattern is additionally overlaid with a blazed grating to separate the generated vortex beam and the zero-order unmodulated light (details in Supplementary Text 4). By cascading LC retarders and 3 metasurfaces with properly designed TCs, the TC carried by the generated vortex beam of the whole structure can be switched among 8 pre-determined integer values.

Experimentally, the TCs carried by the generated vortex beam are characterized by interference fringes with a fundamental Gaussian mode that is aligned to the same polarization state $|V\rangle$. The optical setup is shown in Fig. 2a. First, each single metasurface is characterized with $|H\rangle$ and $|V\rangle$ input, respectively. To characterize a single metasurface, the additional two metasurfaces are removed in Fig. 2a, while the rest elements are fixed. The designed values of TCs and corresponding interference fringes are shown in Fig. 3a. Then, each of the 8 channels are characterized by imposing different phase retardance on the LC retarders. The TCs of 8 channels are expected to be the summation of the TCs of each single metasurface for the picked polarization state. For each channel, the specific combination of polarization states, expected TCs, and the interference fringes are shown in Fig. 3b. As seen in the label of Fig. 3b, the expected TCs of 8 cascaded channels are within the range from *l*=-3 to *l*=+4. To qualitatively observe the TC from the interference fringes, the absolute value of *l* is determined by the number of spiral arms. While the sign of *l* is determined by the spiral direction, clockwise for positive and counter-clockwise for negative. As seen in Fig. 3a and b, the TCs of all generated vortex beams agrees well with the design.

Besides, there can be various TCs of cascaded channels with different settlement of cascading metasurfaces. Another example with the TCs ranging from *l*=-1 to *l*=-8 are experimentally characterized and presented in Supplementary Text 5. Moreover, the interval of TCs can also be varied with non-uniform distribution. A quick example is substituting TCs of the third metasurface with *l*=-3, +3 in Fig. 3a.

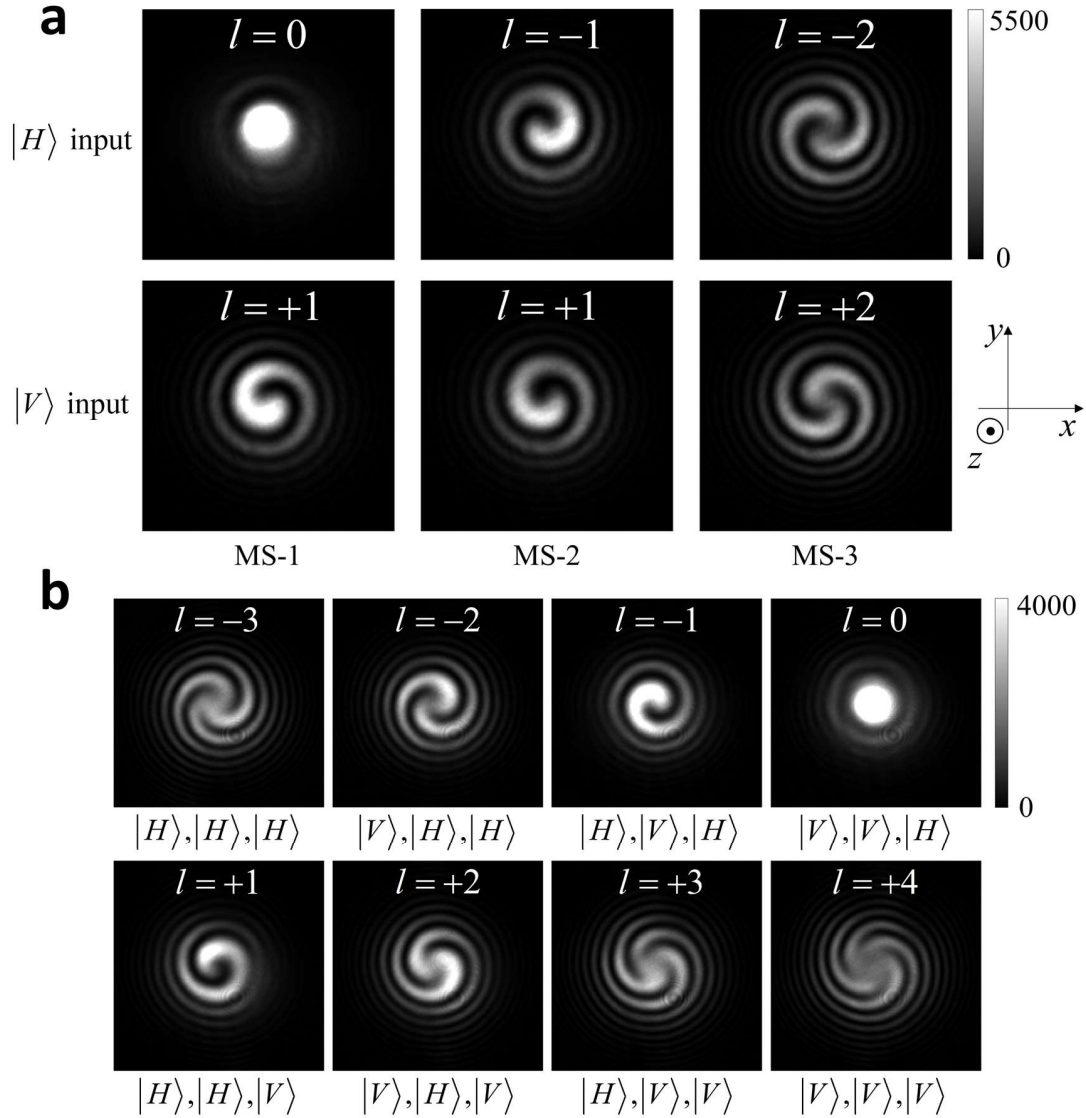

Fig. 3 Experimental results of 8-channel switchable vortex beam generation. The TCs carried by the generated vortex beam are characterized by interference fringes with a fundamental Gaussian mode. (**a**) Each single metasurface with $|H\rangle$ and $|V\rangle$ input. (**b**) 8 channels of cascaded structures, exactly corresponding to 8 combinations of input states of 3 metasurfaces. The input states are labeled in order below the interference fringe image.

**Experimental results of 8-Channel beam steering**

There is only one parameter in aforementioned vortex beam generation. In this section, the functionality of beam steering is further considered as another example with multiple target parameters. In beam steering, the input beam would be deflected. The deflecting direction can be varied in a 2-dimensional plane, which is determined by two parameters of the transverse momenta along $x$ and $y$ axis. To verify our proposal in both 1-dimensional and 2-dimensional parameter space, two cascaded structures (denoted as A and B) for 8-channel tunable beam-steering with different spatial distribution of deflecting directions are designed and characterized. Here, each single metasurface is designed for dual-channel beam steering based on the phase pattern in Eq. 4, which can be also regarded as a polarization beam splitter (PBS). The original designed momenta of each single metasurface and both cascaded structures are listed in Table S1 and Table S2.

The samples are experimentally characterized by the far-field deflecting directions with a vertically incident beam. The optical set up is shown in Fig. 2b. The camera is moved along *z*-axis and the deflected beam spot at different distance is recorded, so that the deflecting direction of $\theta$ and $\varphi$ in spherical coordinates can be calculated. An example of such measurement with $\theta$ and $\varphi$ labeled is shown on the right of Fig. 2b as well. First, the deflecting directions of each single metasurface in cascaded structure A and B are measured and shown in Fig. 4a and Fig. 4c, respectively. As seen in Fig. 4a and Fig. 4c, the first two metasurfaces of both structures are PBSs along *y* axis. The difference between the two structures is introduced only by the third PBS, which is along *y* axis in structure A while along *x* axis in structure B. Then, the deflecting directions of all 8 channels of both cascaded structures are measured and shown in Fig. 4b and Fig. 4d, respectively. As seen in Fig. 4b and Fig. 4d, the deflecting directions of structure A are arranged in an 8×1 line, while those of B are arranged in a 4×2 grid. Furthermore, the angular errors between measured direction and the original design are plotted in Fig. 4e. It can be seen that the maximum error is less than 8'.

To characterize the energy efficiency and cross-talk, the transmittance to each deflecting direction of each channel is measured and plotted in Fig. 4f and Fig. 4g. For structure A, the transmittance to the main deflecting direction is within the range from 15% to 20%, and the cross-talk is below 5% of the input and mainly distributed in adjacent channels. While for structure B, the transmittance is within the range from 20% to 25%, and the cross-talk is more widely distributed. The specific measurement process, original design and raw data can be found in Supplementary Text 6.

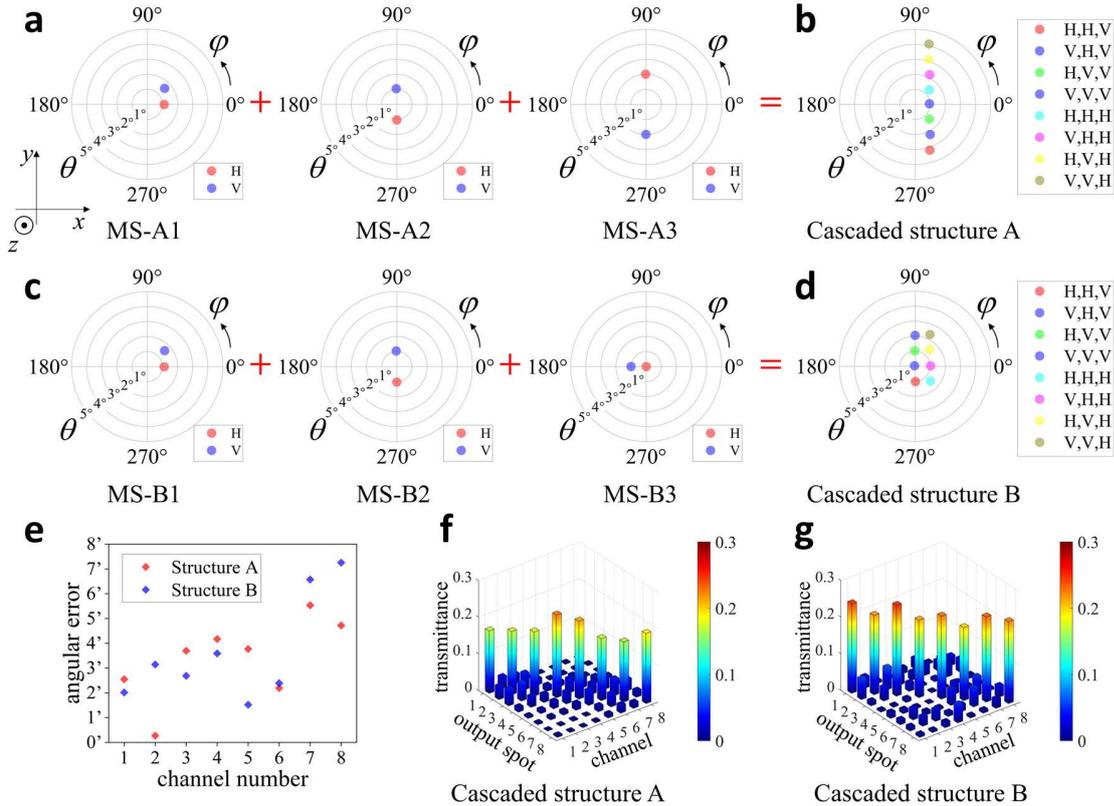

Fig. 4 Experimental results of 8-channel switchable beam steering. (**a**)~(**d**) The deflecting direction of $\theta$ and $\varphi$ in spherical coordinates measured by the deflected beam spot at different distance. (**a**) Each single metasurface in structure A. (**b**) Cascaded structure A. (**c**) Each single metasurface in structure B. (**d**) Cascaded structure B. (**e**) angular errors between measured deflecting direction and the original design. (**f**)(**g**) the transmittance to each deflecting direction of each channel. (**f**) Cascaded structure A. (**g**) Cascaded structure B.

**Discussion**

In the operation principle of our proposal, the whole phase pattern is supposed to be simply the summation of each single phase pattern, without considering the propagation between metasurfaces. In the present work, this is guaranteed by employing a 4-*f* system between each pair of adjacent metasurfaces, so that the input field of the second one can be identical with the output field of the first one (rotated 180° along *z* axis). Actually, such 4-*f* systems are not necessary, since the effect of propagation can be compensated if the distance between metasurfaces is less than several millimeters, with detailed discussions provided in Supplementary Text 7. This is possible by closely cascading metasurfaces and thinner LC retarders as a compact module. Moreover, metasurfaces can be integrated with LCs by similar approach in Ref[49,61]. Thus, it is potential to achieve a fully integrated module while retaining the advantage of low thickness as a single metasurface. Besides, the number of electrodes needed for the electrically switching is determined by LC retarders and almost equal to the layers. Therefore, the design and fabrication of electrode are significantly simpler than pixel-by-pixel electro-optical reconfigurable metasurfaces[27], where the number of electrodes is equal to those of the pixels.

In the present work, the efficiency per layer is 40%~70% and the cross-talk per layer is 3%~5%. As the cascaded layers increases, the stray light, cross-talk and loss would accumulate and become unacceptable. However, they are all non-intrinsic but introduced by the non-ideal factors in metasurface design and fabrication. For example, there would be unmodulated light of metasurface[62–64]. In the vortex beam generation of this work, a blazed grating is overlaid on the phase pattern to spatially separate the unmodulated light[65] and details are presented in Supplementary Text 4. These non-ideal factors can be further reduced with improved design approach of metasurfaces, *e.g.* inverse design[66–70] or adjoint optimization[71,72], thus the accumulated stray light, cross-talk and loss with increased cascading layers can be within control. These methods go beyond the present work, but we have been engaged in related researches.

To conclude, we have achieved $2^N$ electrically switchable channels with $N$ cascaded metasurfaces combined with LC phase retarders. Although the condition is that the phase pattern of cascaded structure should have the same function form with that of a single metasurface, most common dynamic wavefront-controlled functionalities can be covered, including beam steering, vortex beam generation, lens and so on. Moreover, our proposal can be extended to other materials, frequencies and metasurface design approaches. Thus, it is potential to be applied on LiDAR, glasses-free 3D display, OAM (de)multiplexing, and varifocal meta-lens. With more enough channels and richer functionalities, we believe our work would provide a practical way towards metasurfaces with tunable functionalities.

**Acknowledgments**

Funding from the National Key Research and Development Program of China (2023YFB2806703), the National Natural Science Foundation of China (Grant No. U22A6004, 92365210) is greatly acknowledged. This work was also supported by the project of Tsinghua University-Zhuhai Huafa Industrial Share Company Joint Institute for Architecture Optoelectronic Technologies (JIAOT), Beijing National Research Center for Information Science and Technology (BNRist), Frontier Science Center for Quantum Information, Beijing academy of quantum information science, and Tsinghua University Initiative Scientific Research Program. The authors would like to thank Zhe Li, Deyang Kong, and Jiahao Tian for their valuable discussions and helpful comments.

**Author Contributions**

Z.M. and X.F. conceived the idea. Z.M. theoretically verified the principle, designed and performed the simulations, experiments and data analysis. T.T. contributed significantly to the fabrication process. Y.L. contributed to the numerical simulations. Y.L., K.C., F.L, H.S and W.Z. provided useful discussions and comments. Z.M. and X.F. wrote the paper. Y.H. revised the manuscript. All authors approved the manuscript.

Supplementary Materials for

**Electrically switchable $2^N$-channel wave-front control with *N* cascaded polarization-dependent metasurfaces**

Zhiyao Ma, Tian Tian, Yuxuan Liao, Xue Feng*, Yongzhuo Li, Kaiyu Cui, Fang Liu, Hao Sun, Wei Zhang and Yidong Huang*
*Corresponding author: x-feng@tsinghua.edu.cn (X.F.); yidonghuang@tsinghua.edu.cn (Y.H.)

**This PDF file includes:**

Supplementary Text 1 to 6
Figs. S1 to S6
Tables S1 to S2

**Supplementary Text**

1. The Jones matrix of a cascaded structure

For a metasurface with dual-channel phase-only patterns $\varphi_{H,i}(x,y)$ and $\varphi_{V,i}(x,y)$, the Jones matrix of the $i$-th metasurface can be written as:

$$\mathbf{J}_{MS_i}(x,y) = \begin{bmatrix} e^{\varphi_{H,i}(x,y)} & 0 \\ 0 & e^{\varphi_{V,i}(x,y)} \end{bmatrix} \tag{S7}$$

As seen is Eq. S1, the required Jones matrix is a 2×2 diagonal matrix varied with the transverse position $(x,y)$, while the phase of the are determined by the phase-only patterns $\varphi_{H,i}(x,y)$ and $\varphi_{V,i}(x,y)$, and the absolute values of both diagonal elements are 1.

The Jones matrix of each polarization controller is switched between identity matrix or the non-trivial permutation matrix with order 2, which are noted $\mathbf{P}_0$ and $\mathbf{P}_1$ respectively.

$$\mathbf{J}_{PC} = \mathbf{P}_0 = \begin{bmatrix} 1 & 0 \\ 0 & 1 \end{bmatrix} \ or \ \mathbf{J}_{PC} = \mathbf{P}_1 = \begin{bmatrix} 0 & 1 \\ 1 & 0 \end{bmatrix} \tag{S8}$$

Let the profile of input field be $\left|E_{in}\right\rangle = \left|V\right\rangle E_0(x,y)$, where $E_0(x,y)$ is the amplitude distribution and $\left|V\right\rangle = [0,1]^T$ is the polarization state. Then the profile of output field through the whole structure is expressed as:

$$\left|E_{out}\right\rangle = \mathbf{J}_{MS_N}\mathbf{J}_{PC_N}\cdots\mathbf{J}_{MS_2}\mathbf{J}_{PC_2}\mathbf{J}_{MS_1}\mathbf{J}_{PC_1}\left|V\right\rangle E_0(x,y) \tag{S9}$$

Where each of the $\mathbf{J}_{PC}$'s is picked up between $\mathbf{P}_0$ and $\mathbf{P}_1$. It should be mentioned that the cascaded Jones matrices are all unitary, so there would be no loss in theory.

By expanding Eq. S3, it can be seen that the polarization state of $\left|E_{out}\right\rangle$ would be either $\left|V\right\rangle = [0,1]^T$ or $\left|H\right\rangle = [1,0]^T$. And the complex amplitude modulation of $\left|E_{out}\right\rangle$ would be the multiplication of $N$ diagonal element from $N$ metasurfaces, where one of the two diagonal elements is picked up by the incident polarization state at each metasurface. Thus, the whole modulation function of the is also a phase-only pattern $\varphi_\mu(x,y) = \sum_{i=1}^{N} \varphi_{\mu_i,i}(x,y)$, which is mentioned in Eq. 1 of the main text.

2. The experimental setup

The photograph of the experimental set up is shown in Fig. S1. The incident light is a fundamental Gaussian beam collimated from a laser at the wavelength of 1550nm. The polarization state of the incident light is rotated to $|V\rangle$ by a half-wave plate and a polarizer. Then, the radius of the beam incident at the first metasurface is controlled approximately at 200μm by a lens, so that the beam can be covered by the metasurface.

4-*f* systems are employed between each pair of adjacent metasurfaces, so that the input field of the second one can be identical with the output field of the first one (rotated 180° along *z* axis). Then the phase pattern is supposed to be simply the summation of single phase patterns, without considering the propagation between metasurfaces.

The fourth LC retarder is settled after the third metasurface so that the output polarization state of each channel is switched back to $|V\rangle$ and interference with the reference light of $|V\rangle$. The reference light is only used in the characterization of vortex beam generation.

The model of all LC retarders is Thorlabs LCC1223-C. The model of the CCD camera is Hamamatsu InGaAs C12741-03.

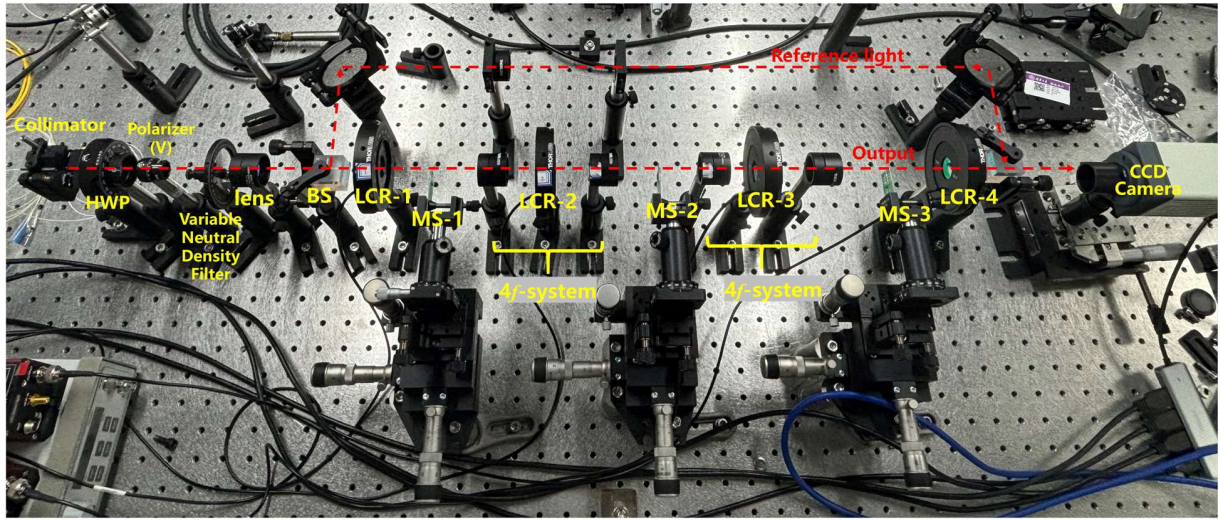

Fig. S1 Photograph of the 8-channel switchable wave-front control with 3-layer setup.

3. Design of the metasurface

First, the amplitude and phase modulation of rectangular nanopillars with different height, period, length and width are numerically calculated by the Finite Difference Time Domain (FDTD) method, with the modeling shown in Fig. S2a. The absolute amplitude modulation values are denoted $T_H, T_V$ for polarization state $|H\rangle$ and $|V\rangle$, while the phase modulation values are denoted $\varphi_H, \varphi_V$. The target is to find a set of rectangular nanopillars with $T_H, T_V$ higher than a threshold, while $(\varphi_H, \varphi_V)$ can fill the region of $[0,2\pi]\times[0,2\pi]$ as dense as possible. So that for any required $(\varphi_H, \varphi_V)$, there would be an optimum nanopillar in the set. Besides, to arrange and fabricate them in a 2-dimensional plane, the height and period should be identical among the set, respectively.

The material of metasurface is considered as amorphous silicon ($\alpha$-Si) on a $SiO_2$ substrate, and the operation wavelength is 1550nm. For the eventually used set, the height is 900nm while the period is 800nm, and the transverse size ranges from 120nm to 700nm. The absolute amplitude modulation values $T_H, T_V$ with respect to the transverse size are shown in Fig. S2b and Fig. S2c, for polarization state $|H\rangle$ and $|V\rangle$, respectively. Actually, Fig. S2b is the mirror-reversal of Fig. S2c along the line $y=x$ according to the symmetry. While the phase modulation values $\varphi_H, \varphi_V$ are shown in Fig. S2d and Fig S2e. Then, the $(\varphi_H, \varphi_V)$'s are scattered in the region of $[0,2\pi]\times[0,2\pi]$ and shown in Fig. S2f. The eventually used set is chosen by the condition that $T_H, T_V$ are both higher than 0.7, and marked by red points. As seen in Fig. S2f, the red points can fill the region of $[0,2\pi]\times[0,2\pi]$ with moderate gaps.

To construct a full metasurface, the target dual-phase pattern should be discretized with the resolution of 800nm, which equals to the period of nanopillars. An example of target phase pattern is seen in Fig. S2g. The target phase pattern for input polarization state $|H\rangle$ is a blazed grating with $(k_x, k_y)=\varepsilon k_0(1,0)$, while that for $|V\rangle$ is a blazed grating with $(k_x, k_y)=\varepsilon k_0(1,1)$, where $\varepsilon=0.01745$. Then, for each pixel, the nanopillar with optimal phase modulation values is chosen from the set. The error of phase modulation values is defined as:

$$\delta = \sqrt{(\varphi_{H,target} - \varphi_{H,nanopillar})^2 + (\varphi_{V,target} - \varphi_{V,nanopillar})^2} \qquad (S10)$$

The nanopillar with minimum error would be chosen. Part of the layout of the constructed metasurface is seen in Fig. S2h. The reconstructed phase pattern is plotted in Fig. S2i according to the simulated phase modulation values of each single nanopillar. The average error is 0.2177rad, while the maximum error is 0.7145rad.

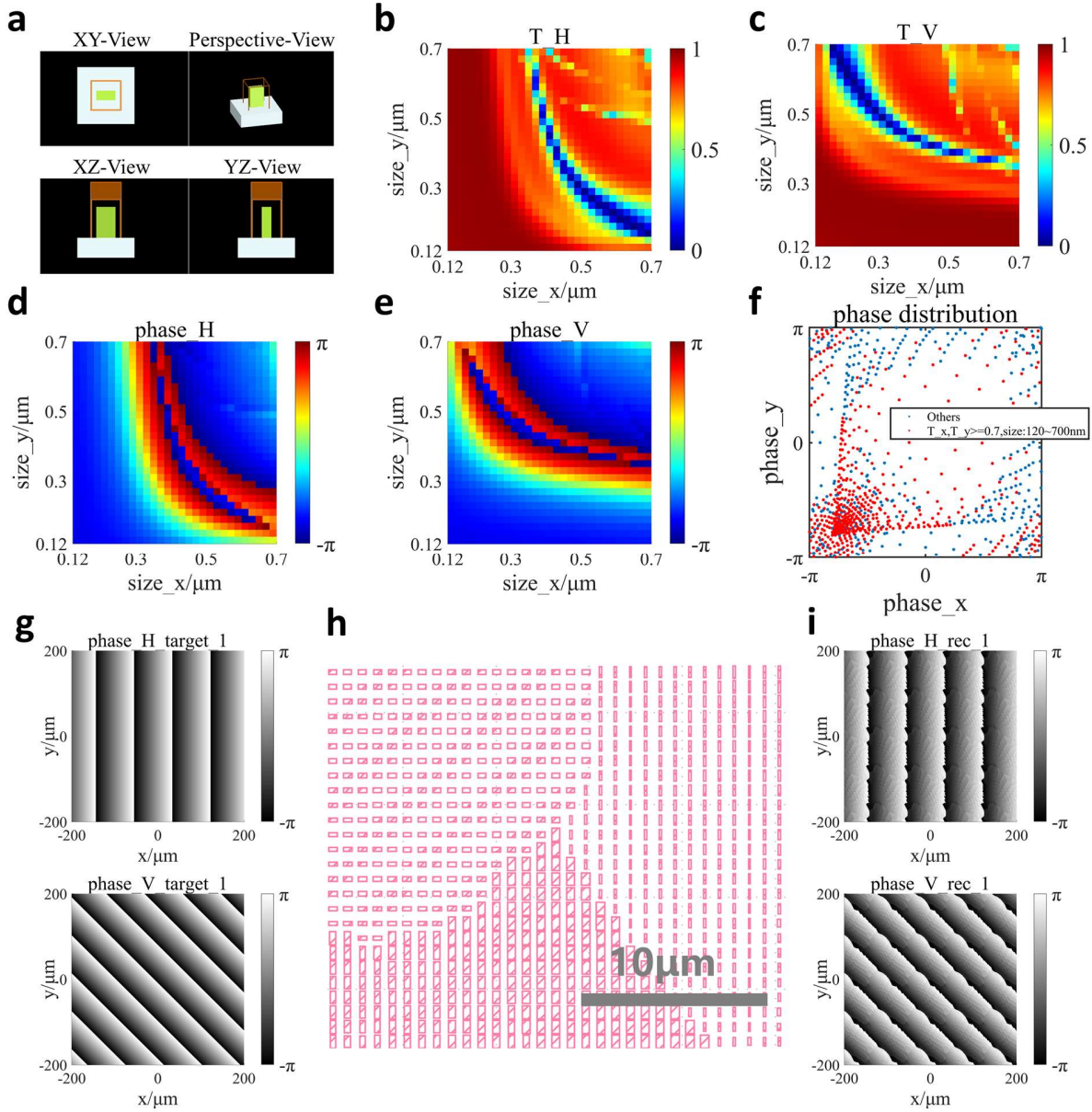

Fig. S2 Design process of the metasurface. (**a**) the modeling of FDTD simulation. (**b**)(**c**) The absolute amplitude modulation values $T_H, T_V$ with respect to the transverse size for polarization state $|H\rangle$ and $|V\rangle$, respectively. (**d**)(**e**) The phase modulation values $\varphi_H, \varphi_V$ with respect to the transverse size for polarization state $|H\rangle$ and $|V\rangle$, respectively. (**f**) $(\varphi_H, \varphi_V)$'s scattered in the region of $[0,2\pi]\times[0,2\pi]$. The eventually used set is chosen by the condition that $T_H, T_V$ are both higher than 0.7, and marked by red points. (**g**)~(**i**) An example of constructing a full metasurface. (**g**) The target phase pattern for input polarization state $|H\rangle$ is a blazed grating with $(k_x, k_y)=\varepsilon k_0(1,0)$, while that for $|V\rangle$ is a blazed grating with $(k_x, k_y)=\varepsilon k_0(1,1)$, where $\varepsilon=0.01745$. (**h**) Part of the layout of the constructed metasurface. (**i**) The reconstructed phase pattern according to the simulated phase modulation values of single nanopillars. The average error is 0.2177rad, while the maximum error is 0.7145rad.

## 4. Blazed grating overlaying method in vortex beam generation

To separate the output vortex beam with the unmodulated light, a blazed grating of $(k_x,k_y)=(0,0.02094k_0)$ is overlaid on the phase pattern of each metasurface for both input polarization states. Taking the three metasurfaces mentioned in Fig. 3a as an example, the target vortex phase pattern with the blazed grating overlaid can be seen in Fig. S3a. Fig. S3b shows the full far-field profiles captured by the CCD camera, where the unmodulated light spots (marked by yellow circles) can be seen below the interference fringes.

Cascading these three metasurfaces would accumulate the deflection introduced by the blazed gratings. Thus, the final far-field profiles in Fig. 3b of the main text contain only the interference fringes. It should be mentioned that the wave-front of unmodulated light of the second and third layer is determined by the input at it, but not always a fundamental Gaussian beam.

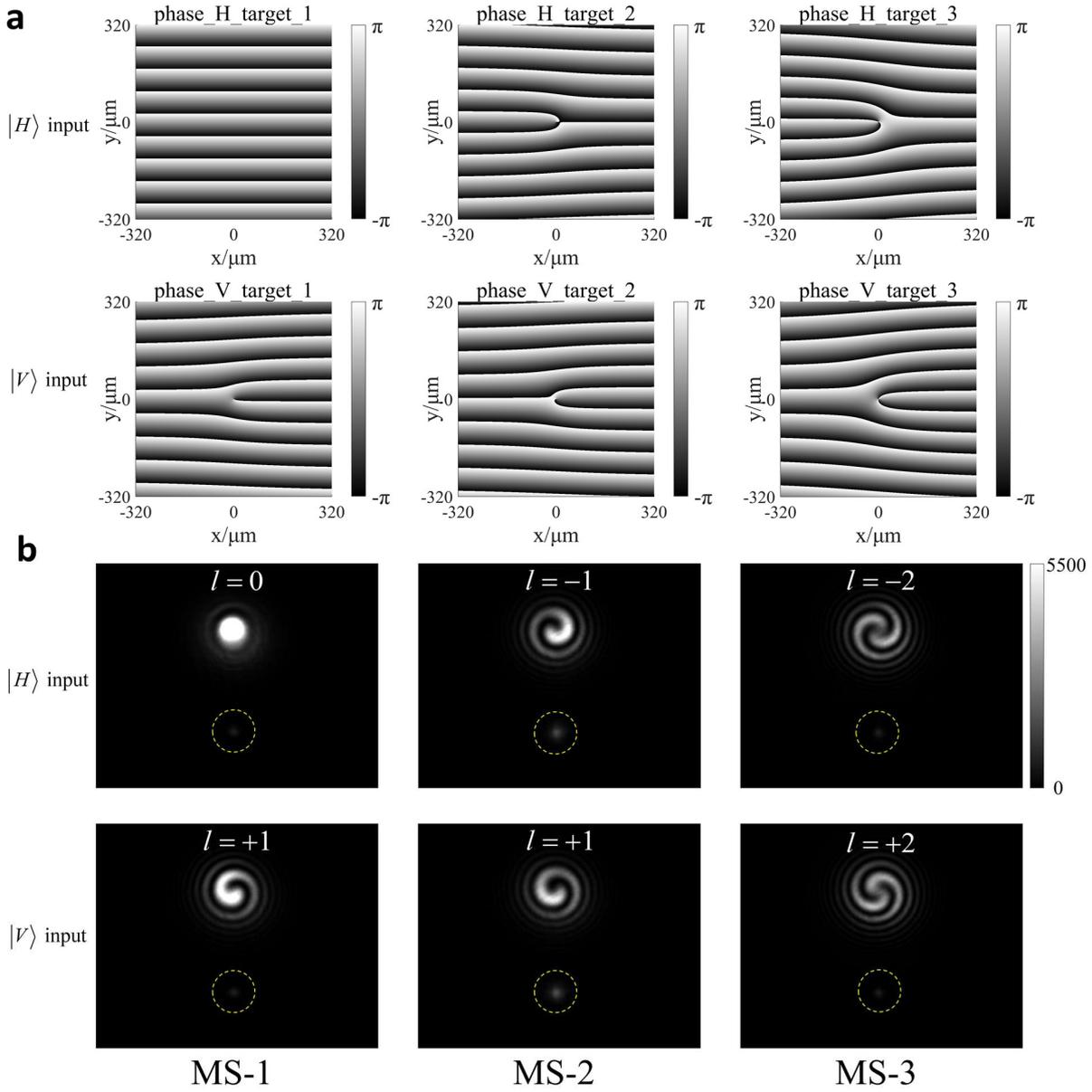

Fig. S3 Example of blazed grating overlaying method in vortex beam generation. The three metasurfaces were mentioned in Fig. 3a. (**a**) The target vortex phase pattern with the blazed grating overlaid. (**b**) The full far-field profiles of Fig. 3a. The gray-level captured by the CCD is rescaled from [0,16383] to [0,5500], which is shown in the color bar. The unmodulated light spots are marked by yellow circles.

## 5. Vortex beam generation with TCs ranging from *l*=-1 to *l*=-8.

In Fig. 3 of the main text, a structure of vortex beam genearation with TCs ranging from *l*=-3 to *l*=+4 are presented. Another example with the TCs ranging from *l*=-1 to *l*=-8 are experimentally characterized as well. The designed TCs interference fringes of each metasurface are shown in Fig. S4a, while the expected TCs and interference fringes of the 8 channels of the cascaded structure are shown in Fig. S4b.

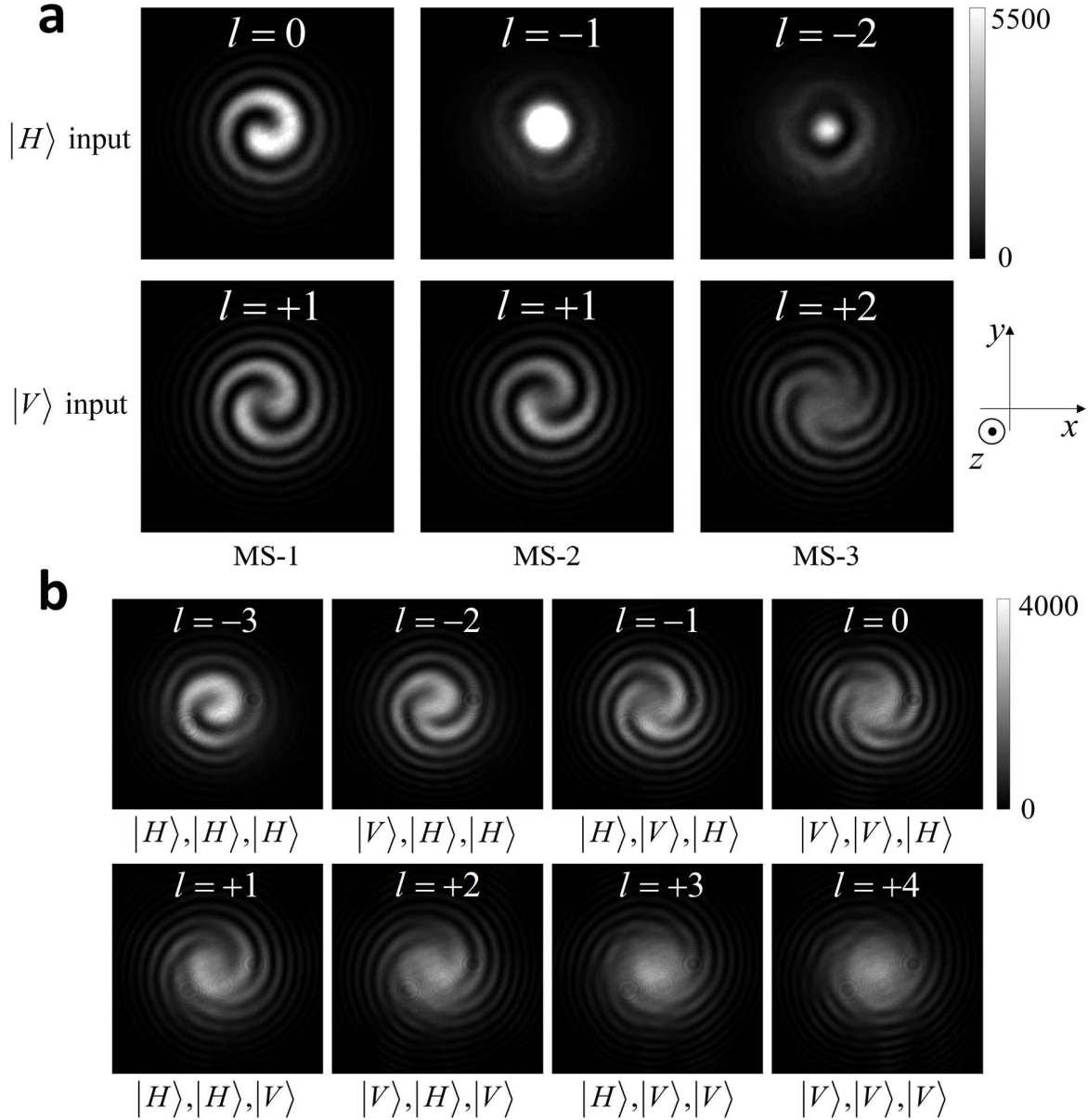

Fig. S4 Experimental results of 8-channel switchable vortex beam generation with TCs ranging from *l*=-1 to *l*=-8. The TCs carried by the generated vortex beam are characterized by interference fringes with a fundamental Gaussian mode. (**a**) Each single metasurface with $|H\rangle$ and $|V\rangle$ input. (**b**) 8 channels of cascaded structures, exactly corresponding to 8 combinations of input states of 3 metasurfaces. The input states are labeled in order below the interference fringe image.

## 6. Characterization of beam steering

Two beam steering structures A and B are mentioned in Fig. 4 of the main text. For structure A, the original designed momenta of each single metasurface and the expected momenta of the 8 channels of the cascaded structure are presented in Table S1. While those for structure B are presented in Table S2.

To characterize the deflecting direction of the output beam in beam steering, the camera is moved along $z$-axis and the far-field profiles at different plane are taken. For each far-field profile, the transverse location of the output beam is obtained by 2-dimensional Gaussian beam fitting method and marked by yellow circles. With the transverse location of the output beam at different $z$'s, the 3-dimensional deflecting direction can be calculated by linear regression. The calculated directions are plotted in Fig. 4a~d of the main text, with the raw data shown in Fig. S5a~c.

To characterize the transmittance of the output beam, $z$ is fixed. The relative intensity of the output beam is obtained by summing the gray-level in the yellow circles. The relative intensity of the input light is obtained the same way in the absence of metasurfaces. Thus, the transmittance can be calculated by the ratio, and shown in Fig. 4f and Fig.4g in the main text.

| Single metasurface | momenta for H input, $(k_{x,H}, k_{y,H})$ | momenta for V input, $(k_{x,V}, k_{y,V})$ |
| --- | --- | --- |
| 1 | $(1,0)\varepsilon k_0$ | $(1,1)\varepsilon k_0$ |
| 2 | $(0,-1)\varepsilon k_0$ | $(0,1)\varepsilon k_0$ |
| 3 | $(0,-2)\varepsilon k_0$ | $(0,2)\varepsilon k_0$ |
| Cascaded channel | momenta $(k_x, k_y)$ | phase retardance of LC retarder |
| H,H,V | $(1,-3)\varepsilon k_0$ | $\pi,0,\pi$ |
| V,H,V | $(1,-2)\varepsilon k_0$ | $0,\pi,\pi$ |
| H,V,V | $(1,-1)\varepsilon k_0$ | $\pi,\pi,0$ |
| V,V,V | $(1,0)\varepsilon k_0$ | $0,0,0$ |
| H,H,H | $(1,1)\varepsilon k_0$ | $\pi,0,0$ |
| V,H,H | $(1,2)\varepsilon k_0$ | $0,\pi,0$ |
| H,V,H | $(1,3)\varepsilon k_0$ | $\pi,\pi,\pi$ |
| V,V,H | $(1,4)\varepsilon k_0$ | $0,0,\pi$ |

Table S1 Original design of 8-channel switchable beam steering structure A. $\varepsilon=0.01745$. The momenta of cascaded channels are calculated by summing the momenta for the corresponding input polarization state of each single metasurface. The phase retardance given in the table is determined by whether to change the polarization states between adjacent metasurfaces.

| Single metasurface | momenta for H input, $(k_{x,H}, k_{y,H})$ | momenta for V input, $(k_{x,V}, k_{y,V})$ |
|---|---|---|
| 1 | $(1,0)\varepsilon k_0$ | $(1,1)\varepsilon k_0$ |
| 2 | $(0,-1)\varepsilon k_0$ | $(0,1)\varepsilon k_0$ |
| 3 | $(0,0)\varepsilon k_0$ | $(-1,0)\varepsilon k_0$ |
| **Cascaded channel** | **momenta $(k_x, k_y)$** | **phase retardance of LC retarder** |
| H,H,V | $(0,-1)\varepsilon k_0$ | $\pi,0,\pi$ |
| V,H,V | $(0,0)\varepsilon k_0$ | $0,\pi,\pi$ |
| H,V,V | $(0,1)\varepsilon k_0$ | $\pi,\pi,0$ |
| V,V,V | $(0,2)\varepsilon k_0$ | $0,0,0$ |
| H,H,H | $(1,-1)\varepsilon k_0$ | $\pi,0,0$ |
| V,H,H | $(1,0)\varepsilon k_0$ | $0,\pi,0$ |
| H,V,H | $(1,1)\varepsilon k_0$ | $\pi,\pi,\pi$ |
| V,V,H | $(1,2)\varepsilon k_0$ | $0,0,\pi$ |

Table S2 Original design of 8-channel switchable beam steering structure B. $\varepsilon=0.01745$.

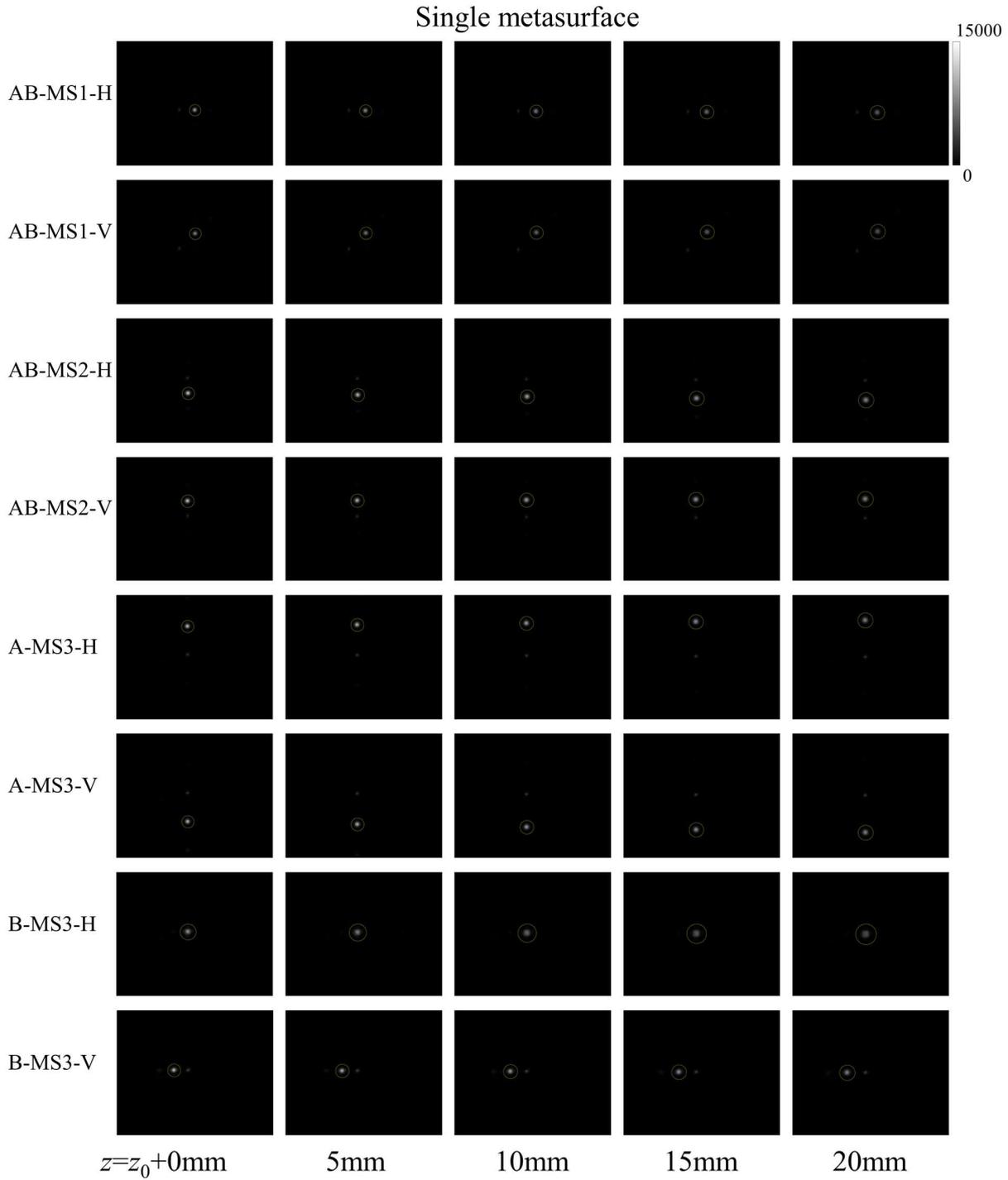

Fig. S5a The raw data of each single metasurface in Fig. 4a and Fig 4c. The first metasurface is identical of both structure A and structure B (denoted as AB-MS1), as well as the second metasurface (denoted as AB-MS2). The third metasurface of structure A is denoted as A-MS3, while the third metasurface of structure B is denoted as B-MS3. The fitted output beams are marked in yellow circles.

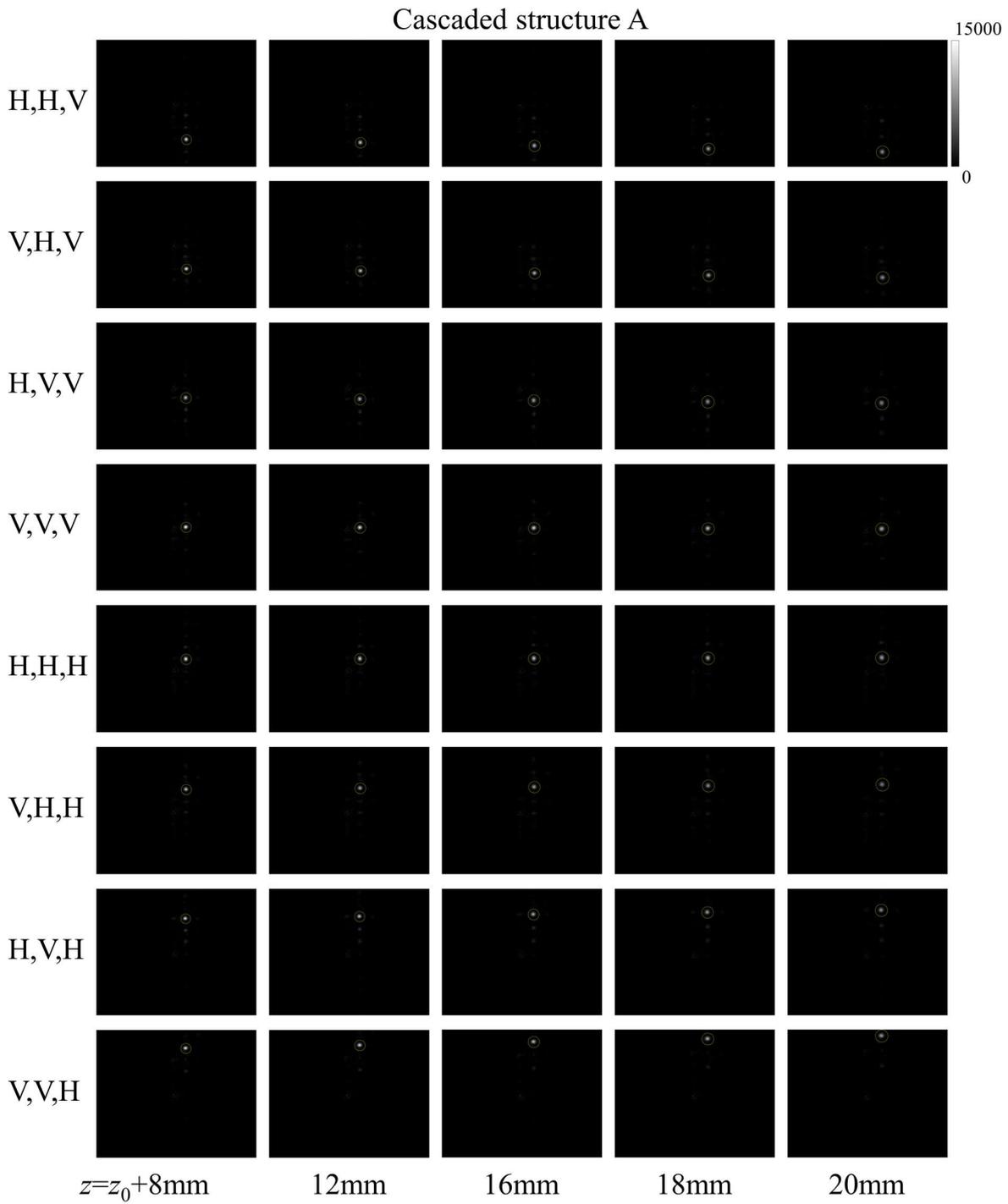

Fig. S5b The raw data of cascaded structure A in Fig. 4b.

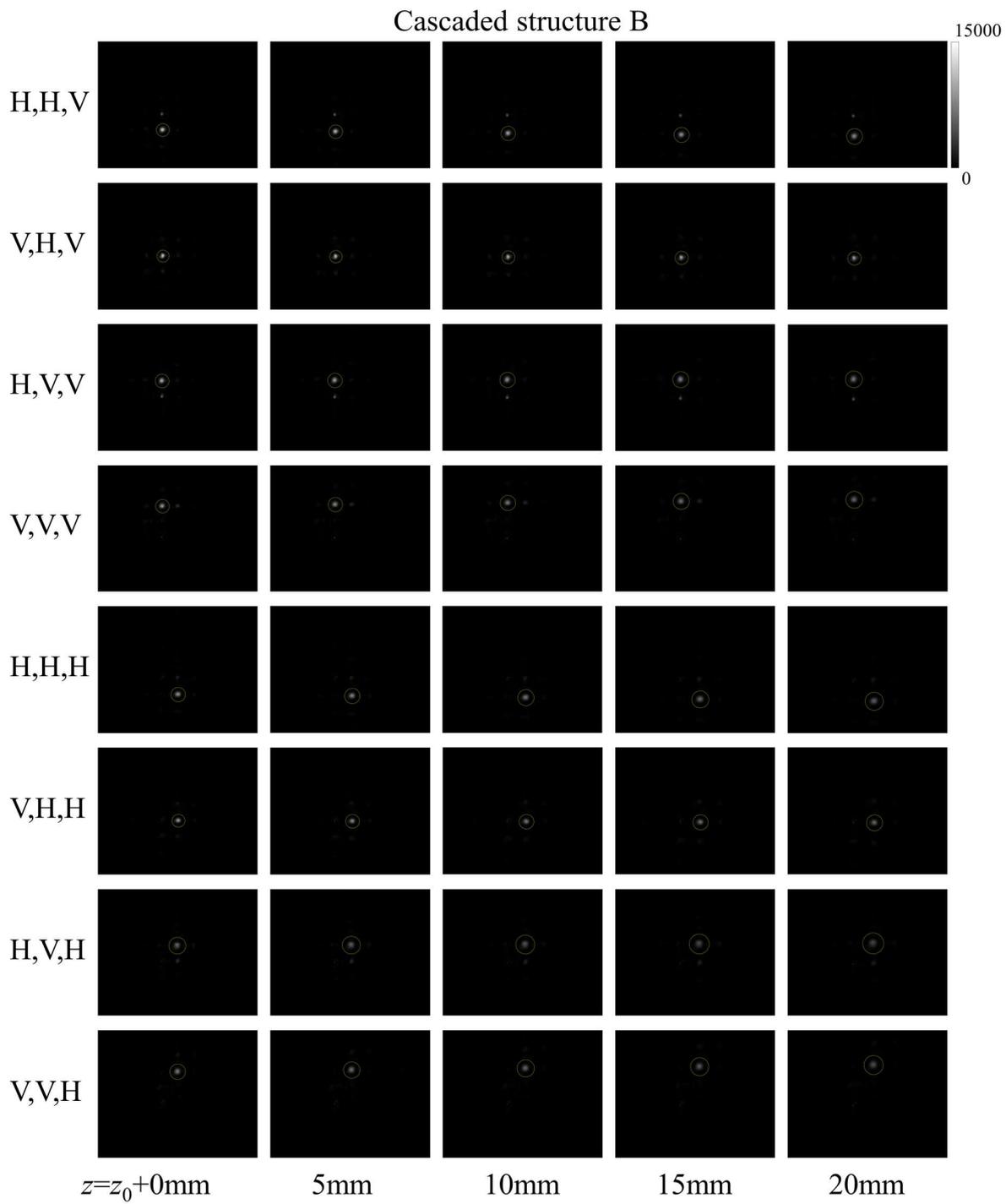

Fig. S5c The raw data of cascaded structure A in Fig. 4d.

## 7. Discussions about implementing compact module

For a compact module with closely cascaded metasurfaces and LC retarders (without 4-*f* systems), the effect of propagation should be considered in the calculation of the whole phase pattern.

For functionalities of vortex beam generation, the propagation effect is only the divergence of beams. If the distance of propagation is not too long, the divergence of beams can be compensated by overlaying a lens pattern with appropriate focus length on the metasurface. Suppose the width *w* of incident beam at each metasurface is constant, the maximum distance between adjacent metasurface can be calculated by the evolution of Gaussian beams through propagation:

$$d \leq \frac{\pi w^2}{\lambda} \tag{S11}$$

For example, suppose *w*=200μm, the maximum distance between adjacent metasurface is 81.1mm. This distance would be far enough for cascading metasurfaces and LC retarders. According to the distance between adjacent metasurfaces, the required focus length of overlaid lens pattern can be calculated by the evolution of Gaussian beams through propagation as well. Examples with *w*=100μm, 200μm, 300μm are shown in Fig. S6a.

The similar approach can be applied for varifocal lens. Actually, the equivalent focal length of the cascaded structure can be calculated by considering the propagation, thus the effect of propagation on the whole focal lengths can be compensated by inversely adjusting the focal length of each single metasurface.

Besides the divergence of beams, there would be another problem in beam steering. Since the beam is deflected, it is possible to go beyond the region of next metasurface through propagation. The condition to avoid such problem is accessible by geometrical optics shown in Fig. S6b:

$$d \sum_{i=1}^{N-1} \tan \theta_i < r_{MS} - w \tag{S12}$$

Where *d* is the distance between metasurfaces, $\theta_i$ is the angle between the optical axis and the deflection direction of the *i*-th metasurface, $r_{MS}$ is the radius of metasurface, and *w* is the radius of the beam (supposing the lens-overlaying method is applied so that the radius of beam at each metasurface is constant). Such condition is highly possible to be satisfied. For example, suppose $r_{MS}$=500μm, *w*=200μm and the maximum deflection angle is 5°. Then, Eq. S6 is guaranteed if (*N*-1)*d*<3.43mm. Since the thickness of metasurface with substrate is approximately 300μm, and the thickness of an integrated full-wave LC retarder can be less than 10μm (Ref[60]), 3.43mm would be enough for even 10 layers, which corresponds to ~$10^3$ ($2^{10}$) channels.

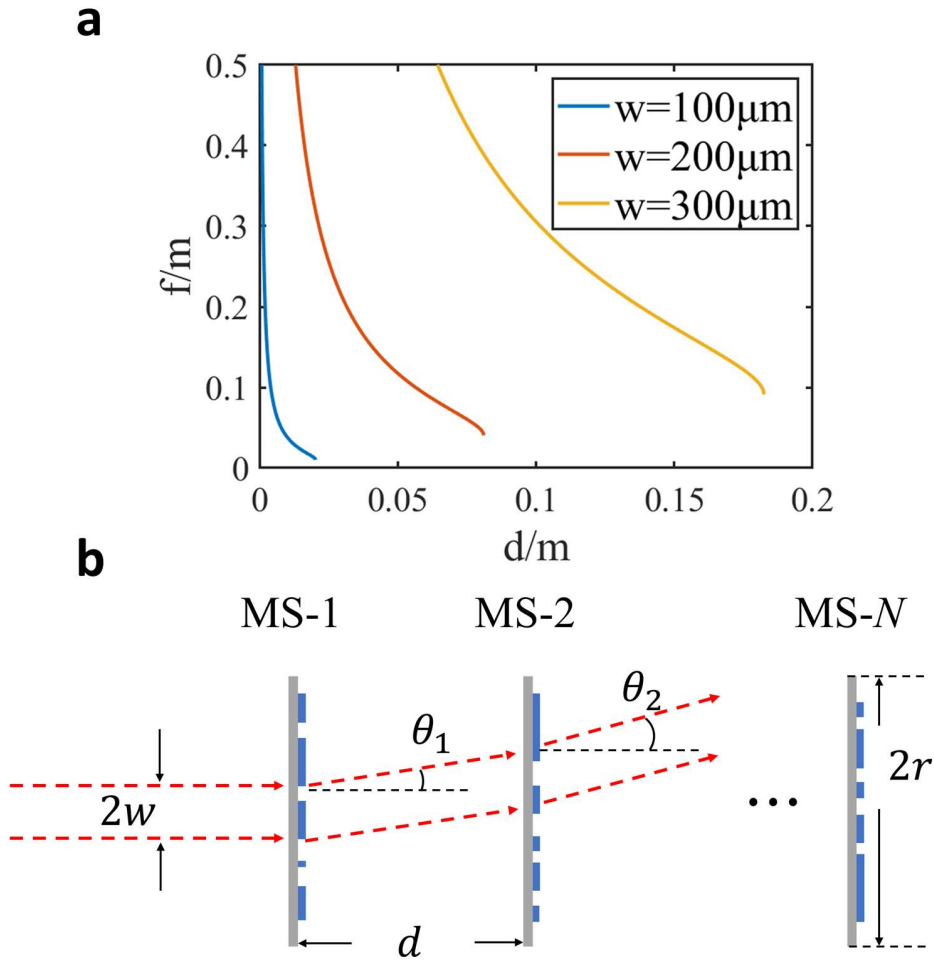

Fig. S6 Discussions of a compact module without 4-*f* systems. (**a**) For functionalities of vortex beam generation, the divergence of beams can be compensated by overlaying a lens pattern with appropriate focus length on the metasurface. The required focus length of overlaid lens pattern with respect to to the distance between adjacent metasurfaces. (**b**) For functionalities of beam steering, the condition to avoid the beam missing the region of metasurface through propagation is accessible by geometrical optics and seen in Eq. S6.